\documentclass[fleqn,usenatbib,useAMS]{mnras}
 \usepackage{graphicx}
 \usepackage{amsmath}
 \usepackage{amssymb}
 \usepackage[T1]{fontenc}
 \usepackage{ae,aecompl}
 \usepackage{txfonts}
 \usepackage{enumerate}

 \title[Orbital evolution of Atira asteroid 2019~AQ$_{3}$]
       {Understanding the evolution of Atira-class asteroid 2019~AQ$_\mathbf{3}$, 
        a major step towards the future discovery of the Vatira population} 

 \author[C. de la Fuente Marcos and R. de la Fuente Marcos]
        {C.~de~la~Fuente~Marcos$^{1}$\thanks{E-mail: nbplanet@ucm.es}
         and
         R. de la Fuente Marcos$^{2}$ \\
         $^1$ Universidad Complutense de Madrid,
              Ciudad Universitaria, E-28040 Madrid, Spain \\
         $^2$AEGORA Research Group,
             Facultad de Ciencias Matem\'aticas,
             Universidad Complutense de Madrid,
             Ciudad Universitaria, E-28040 Madrid, Spain}
 \date{Accepted 2019 May 21.
       Received 2019 May 20;
       in original form 2019 January 10}
 \pubyear{2019}
 
\begin{document}
  \label{firstpage}
  \pagerange{\pageref{firstpage}--\pageref{lastpage}}
  \maketitle

  \begin{abstract}
     Orbiting the Sun at an average distance of 0.59~au and with the shortest 
     aphelion of any known minor body, at 0.77~au, the Atira-class asteroid 
     2019~AQ$_{3}$ may be an orbital outlier or perhaps an early indication 
     of the presence of a new population of objects: those following orbits 
     entirely encompassed within that of Venus, the so-called Vatiras. Here, 
     we explore the orbital evolution of 2019~AQ$_{3}$ within the context of 
     the known Atiras to show that, like many of them, it displays a 
     present-day conspicuous coupled oscillation of the values of 
     eccentricity and inclination, but no libration of the value of the 
     argument of perihelion with respect to the invariable plane of the Solar  
     system. The observed dynamics is consistent with being the result of the 
     combined action of two dominant perturbers, the Earth--Moon system and 
     Jupiter, and a secondary one, Venus. Such a multiperturber-induced 
     secular dynamics translates into a chaotic evolution that can eventually 
     lead to a resonant behaviour of the Lidov--Kozai type. Asteroid 
     2019~AQ$_{3}$ may have experienced brief stints as a Vatira in the 
     relatively recent past and it may become a true Vatira in the future, 
     outlining possible dynamical pathways that may transform Atiras into 
     Vatiras and vice versa. Our results strongly suggest that 2019~AQ$_{3}$ 
     is only the tip of the iceberg: a likely numerous population of similar 
     bodies may remain hidden in plain sight, permanently confined inside the 
     Sun's glare. 
  \end{abstract}

  \begin{keywords}
     methods: numerical -- methods: statistical -- celestial mechanics --
     minor planets, asteroids: general -- 
     minor planets, asteroids: individual: 2019~AQ$_{3}$ --
     planets and satellites: individual: Venus. 
  \end{keywords}

  \section{Introduction}
     The innermost section of the Solar system, that within Venus' orbit, is still far from well studied as it mostly remains confined 
     inside the Sun's glare as seen from the Earth. So far, no minor bodies have ever been observed to orbit the Sun entirely within Venus' 
     path, but they are assumed to exist: the so-called Vatiras \citep{2012Icar..217..355G}. Bordering the hypothetical Vatira orbital realm 
     (aphelion distance, $Q$, in the range 0.307--0.718~au) there is only one known Atira, 2019~AQ$_{3}$ with $Q=0.77$~au. 

     Atira-class asteroids ---also known in the literature as Apoheles \citep{1998MPBu...25...42T},\footnote{\url{http://defendgaia.org/bobk/ccc/cc070998.html}}
     or Inner-Earth objects, IEOs \citep{2000Icar..143..421M,2008Icar..198..284Z}--- have 0.718~au$<Q<$ 0.983~au (Greenstreet et al. 
     \citeyear{2012Icar..217..355G}). As of 2019 May 20, there are only 19 known Atiras because, like Venus or Mercury, they can only be 
     observed near their maximum elongation (east or west; see e.g. \citealt{2003Icar..163..389M}), when their angular separation from the 
     Sun is greatest ---in the interval (18\degr, 28\degr) for Mercury and (45\degr, 47\degr) for Venus. The relevance and implications of 
     studying minor bodies at small solar elongations was first discussed by \citet{1998DPS....30.1604T} and \citet{1998DPS....30.1603W}, 
     and the origin of those known remains an open question \citep{2010DPS....42.1309G}. Although a long-term stable asteroid belt located 
     at 0.09--0.21~au from the Sun has been predicted \citep{1999Natur.399...41E,2002MNRAS.333L...1E}, no positive detections have been 
     reported yet \citep{2000Icar..148..312D,2001A&A...368.1108S,2013Icar..223...48S}. It is however unclear whether other processes may 
     remove material from this region (see e.g. \citealt{1994Natur.371..315F,2000Icar..143..360S,2000Icar..148..147V,2003A&A...411..291W}). 

     Atira-class asteroids can experience close encounters with Mercury, Venus, and the Earth--Moon system, but \citet{2018RNAAS...2b..46D} 
     have shown that many known Atiras display a conspicuous behaviour that somehow resembles the one induced by the Lidov--Kozai mechanism 
     \citep{1962AJ.....67..591K,1962P&SS....9..719L}; such a circumstance opens the door to a chaotic but relatively stable dynamical 
     scenario. This interpretation is consistent with the one presented by \citet{2016MNRAS.458.4471R}, where it is argued that many Atiras 
     remain on regular orbits for at least 1~Myr. Here, we investigate the dynamical evolution of 2019~AQ$_{3}$, a recently discovered 
     near-Earth asteroid (NEA) that has the shortest aphelion of any known minor body, 0.77~au. This paper is organized as follows. In 
     Section~2, we present the available data on this object and the tools used in our study. The orbital evolution of 2019~AQ$_{3}$ is 
     explored in Section~3. Our results are discussed in Section~4 and our conclusions are summarized in Section~5. 

  \section{Data and methods}
     The source of most of the data used in this research is Jet Propulsion Laboratory's Solar System Dynamics Group Small-Body Database 
     (JPL's SSDG SBDB, \citealt{2001DPS....33.5813G,2011jsrs.conf...87G,2015IAUGA..2256293G})\footnote{\url{https://ssd.jpl.nasa.gov/sbdb.cgi}}
     and JPL's \textsc{horizons}\footnote{\url{https://ssd.jpl.nasa.gov/?horizons}} ephemeris system \citep{1996DPS....28.2504G,ST98,GY99}. 
     This includes orbit determinations, covariance matrices, initial conditions (positions and velocities in the barycentre of the Solar 
     system) for planets and minor bodies referred to epoch JD 2458600.5 (2019-Apr-27.0) TDB (Barycentric Dynamical Time) ---which is the 
     zero instant of time in the figures, J2000.0 ecliptic and equinox--- ephemerides, and other input data. When quoted, statistical 
     parameters have been computed in the usual way (see e.g. \citealt{2012psa..book.....W}).

     \subsection{Atira-class asteroid 2019~AQ$_\mathbf{3}$: the data}
        Asteroid 2019~AQ$_{3}$ was discovered on 2019 January 4 \citep{2019MPEC....A...88B} by the Zwicky Transient Facility observing 
        system \citep{2014SPIE.9147E..79S, 2017NatAs...1E..71B} at Palomar Mountain. The Pan-STARRS project \citep{2004SPIE.5489...11K,
        2004AAS...204.9701K} team found precovery images acquired in 2015 \citep{2019MPEC....A...97C}. With both old and new data, the orbit 
        determination of 2019~AQ$_{3}$ shown in Table~\ref{elements} was computed on 2019 February 3 and it is based on 49 observations for 
        a data-arc span of 1199~d. 

        Asteroid 2019~AQ$_{3}$ is relatively large with an absolute magnitude of 17.6~mag (assumed $G=0.15$), which suggests a diameter in 
        the range 0.52--4.00~km for an assumed albedo in the range 0.60--0.01. It is a member of the Atira dynamical class (see above), but 
        also an NEA because its perihelion distance is under 1.3~au. Its aphelion distance, 0.77~au, is the shortest of any known minor body 
        ---followed by (418265) 2008~EA$_{32}$ at 0.80~au. It has the shortest sidereal orbital period of any known Atira and the second 
        shortest of any known minor body with nearly 165~d; Aten-class ($a<1$~au and $Q>0.983$~au) asteroid 2016~XK$_{24}$ has the shortest, 
        with slightly over 161~d, but its orbit determination is very uncertain as it is based on seven observations for a data-arc span of 
        1~d. Within the Atira group, 2019~AQ$_{3}$ has the highest value of the orbital inclination, 47\fdg2. It is probably the third 
        largest member of the group, together with 2018~JB$_{3}$; the synchronous binary asteroid (163693)~Atira \citep{2017CBET.4347....1R} 
        is the largest with 4.8~km, followed by 418265.
%
%
     \begin{table}
      \centering
      \fontsize{8}{11pt}\selectfont
      \tabcolsep 0.15truecm
      \caption{Values of the Heliocentric Keplerian orbital elements of 2019~AQ$_{3}$ and their associated 1$\sigma$ uncertainties. The 
               orbit determination is referred to epoch JD 2458600.5 (2019-Apr-27.0) TDB (J2000.0 ecliptic and equinox). Source: JPL's SBDB.
              }
      \begin{tabular}{lcccc}
       \hline
        Orbital parameter                                 &   & Value$\pm$1$\sigma$ uncertainty \\
       \hline
        Semimajor axis, $a$ (au)                          & = &   0.58866153$\pm$0.00000008 \\
        Eccentricity, $e$                                 & = &   0.3143098$\pm$0.0000005   \\
        Inclination, $i$ (\degr)                          & = &  47.2186$\pm$0.0009         \\
        Longitude of the ascending node, $\Omega$ (\degr) & = &  64.4873$\pm$0.0004         \\
        Argument of perihelion, $\omega$ (\degr)          & = & 163.1518$\pm$0.0005         \\
        Mean anomaly, $M$ (\degr)                         & = & 118.4994$\pm$0.0009         \\
        Perihelion, $q$ (au)                              & = &   0.4036395$\pm$0.0000003   \\
        Aphelion, $Q$ (au)                                & = &   0.77368361$\pm$0.00000010 \\
        Absolute magnitude, $H$ (mag)                     & = &  17.6$\pm$0.5               \\
       \hline
      \end{tabular}
      \label{elements}
     \end{table}
%
%

     \subsection{$\bmath{N}$-body simulations}
        In order to explore the orbital evolution of 2019~AQ$_{3}$, we have performed full $N$-body simulations using a 
        software\footnote{\url{http://www.ast.cam.ac.uk/~sverre/web/pages/nbody.htm}} that implements a fourth-order version of the Hermite 
        integration scheme \citep{1991ApJ...369..200M,2003gnbs.book.....A}. We have carried out calculations under the Newtonian 
        approximation as described by \citet{2012MNRAS.427..728D}, but also under the post-Newtonian approximation as in 
        \citet{2015MNRAS.446.1867D} because the perihelion distance of 2019~AQ$_{3}$ becomes comparable (see Section~\ref{relcal}) to those 
        of the so-called relativistic asteroids (see table~2 in \citealt{2008CeMDA.101..289B}). The impact of the uncertainties of the orbit
        determination on our results has been evaluated using the covariance matrix methodology described in \citet{2015MNRAS.453.1288D}.

        The integrator used in this work is not symplectic; therefore, it can be argued that the accumulation of errors during long 
        calculations could be important and, since the trajectory of 2019~AQ$_{3}$ and those of other Atiras studied here are chaotic, this 
        may lead to significant changes in the long-term evolution of these objects as portrayed by our simulations. The issue of symplectic
        integration versus other approaches, such as the fourth-order version of the Hermite integration scheme \citep{1991ApJ...369..200M}
        used here or time-symmetric versions of these algorithms, has been discussed in detail by e.g. \citet{2015MNRAS.452.1934H,
        2018MNRAS.475.5570H} and it is outside the scope of this work. It is however worth to mention that \citet{2012MNRAS.427..728D} show 
        explicitly (see their fig.~3) that results from the integration scheme used here are consistent with those obtained using other 
        techniques, e.g. \citet{2003ApJ...592..620V} or \citet{2011A&A...532A..89L}.

  \section{Orbital evolution}
     The analysis in \citet{2018RNAAS...2b..46D} pointed out that many Atiras have arguments of perihelion, $\omega$, in the neighbourhood 
     of 0{\degr} or 180{\degr}, which means that the nodal points ---where the orbit crosses the ecliptic--- are located at perihelion and 
     at aphelion \citep{1989Icar...78..212M}. However, this could be due to observational bias (see e.g. \citealt{2016MNRAS.458.4471R}) as 
     the Atiras are preferentially discovered near aphelion. 

     For 2019~AQ$_{3}$, $\omega=163\fdg2$ and given the fact that $Q=0.77$~au (the semimajor axis of Venus is 0.72~au with a nearly circular 
     orbit) the direct perturbation of Venus might be significant. On the other hand, \citet{2018RNAAS...2b..46D} also discussed an 
     eccentricity--inclination ($e-i$) inverse relationship that is ubiquitous within the Atira group; this property is also present in the 
     case of 2019~AQ$_{3}$, relatively low eccentricity (0.31) but high inclination (47\fdg2), see Table~\ref{elements}. These facts suggest 
     that the Lidov--Kozai mechanism might control the orbital evolution of these objects; the Lidov--Kozai resonance, within the context of 
     NEAs, has been studied by e.g. \citet{1996A&A...307..310M} and \citet{2015A&A...580A.109D}. The Lidov--Kozai mechanism in the context of 
     multiplanet systems has been studied by \citet{2009A&A...493..677L}. However, the Lidov--Kozai mechanism requires the concurrent 
     oscillation of the values of three orbital parameters (in the frame of reference of the invariable plane of the studied system): $e$, 
     $i$, and $\omega$. Direct $N$-body simulations should be able to either confirm or reject the presence of a simultaneous, repetitive 
     variation of the values of $e$, $i$, and $\omega$ throughout the orbital evolution of the Atiras.
%
%
     \begin{figure*}
       \centering
        \includegraphics[width=0.49\linewidth]{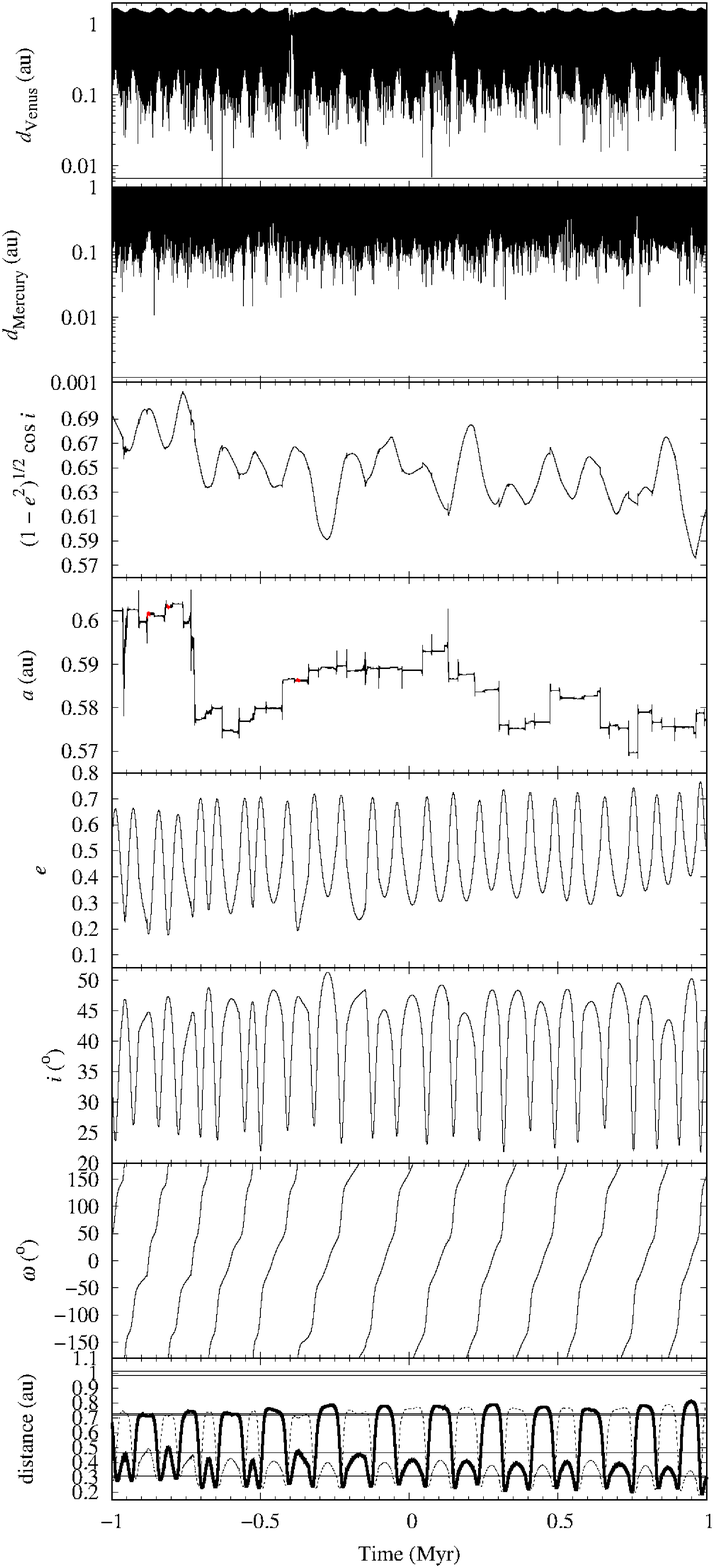}
        \includegraphics[width=0.495\linewidth]{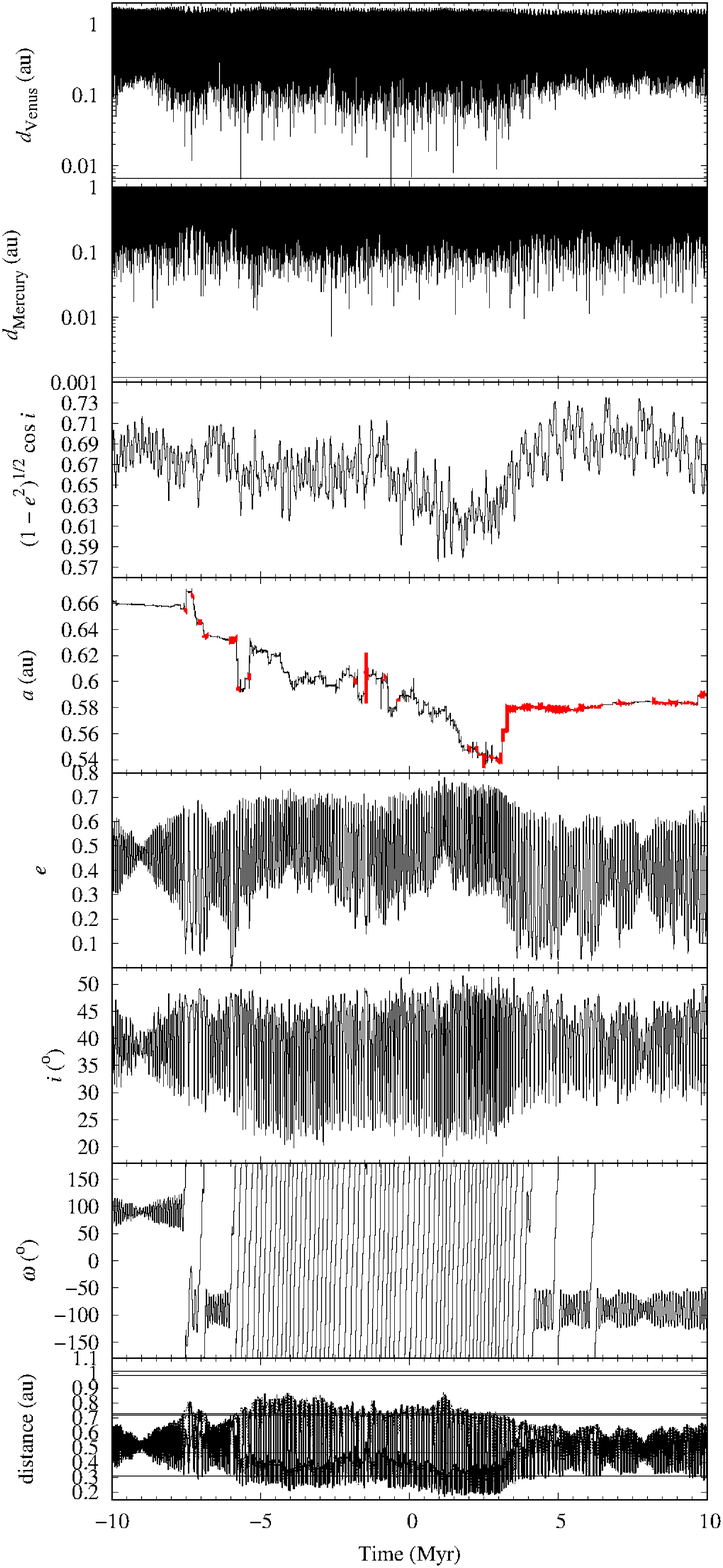}
        \caption{Evolution of the values of relevant parameters for the nominal orbit of 2019~AQ$_{3}$ (the left-hand side set of panels is 
                 a magnified version of the one on the right-hand side): distance from Venus (top panel, with Hill radius of Venus, 
                 0.0067~au), distance from Mercury (second to top panel, with Hill radius of Mercury, 0.0012~au), value of the Lidov--Kozai 
                 parameter (third to top panel), semimajor axis (fourth to top panel, when the object becomes a Vatira it is plotted in 
                 thick red/grey), eccentricity (fourth to bottom panel), inclination (third to bottom panel), argument of perihelion (second 
                 to bottom panel), and nodal distances from the Sun (bottom panel, to descending node as thick line and to ascending node as 
                 dotted line, the aphelion and perihelion distances of Mercury, Venus, and the Earth are indicated as well).
                }
        \label{evolution}
     \end{figure*}
%
%

     \subsection{Short-term evolution}
        Figure~\ref{evolution}, left-hand side set of panels, shows the short-term evolution (from 1~Myr into the past to 1~Myr into the 
        future) of the values of a number of relevant parameters for the nominal orbit of 2019~AQ$_{3}$. Close encounters with the 
        Earth--Moon system are not possible, but both the top and second to top panels in Fig.~\ref{evolution} show that the peculiar 
        dynamics of 2019~AQ$_{3}$ leads to avoiding close encounters with both Venus and Mercury for the most part; only a few flybys within 
        the Hill radius of Venus, 0.0067~au, are recorded in our calculations. The third to top panel shows the evolution of the value of 
        the so-called Lidov--Kozai parameter, $\sqrt{1-e^{2}}\ \cos{i}$ that oscillates significantly during most of the displayed time 
        interval. \citet{1996A&A...307..310M} and \citet{2015A&A...580A.109D} have shown that the value of the Lidov--Kozai parameter of NEAs 
        tends to change only slightly when the Lidov--Kozai resonance is in effect (see also Section~\ref{LKev}).

        The value of the semimajor axis (fourth to top panel) changes by about 5 per cent during the time interval displayed, drifting from 
        one quasi-constant value to a relatively close one (also quasi-constant) for most of the integrated time. The evolution is chaotic 
        as orbits starting arbitrarily close to each other diverge relatively quickly (not shown in the figures), but the overall dynamics 
        resembles that of stable chaotic orbits (see e.g. \citealt{1992Natur.357..569M}). On the other hand, the values of $e$ (fourth to 
        bottom panel) and $i$ (third to bottom panel) oscillate, alternating high $e$ and $i$. It is clear that, although the evolution 
        somehow resembles the one found in the Lidov--Kozai scenario for $e$ and $i$ \citep{1962AJ.....67..591K,1962P&SS....9..719L}, the 
        overall dynamical situation is different as the value of $\omega$ (second to bottom panel) circulates. The lack of libration in 
        $\omega$ confirms that this object is not currently subjected to a Lidov--Kozai resonance (but see Section~\ref{invar} for a proper 
        analysis in the frame of reference of the invariable plane of the system).

        The nodal distances from the Sun (bottom panel) show the location of the points where the orbit crosses the ecliptic; encounters
        with planetary bodies may take place there. An oscillation is observed, but the current layout ($t=0$) shows the descending node
        located between Mercury's perihelion and aphelion, and the ascending one beyond Venus. This explains well why the distances from
        Mercury and Venus remain large enough to avoid close flybys; the dynamical context that controls the orbital evolution of
        2019~AQ$_{3}$ is protecting it from approaching the innermost planets too closely. A similar behaviour has been previously observed
        in the case of Venus co-orbitals (see e.g. \citealt{2012MNRAS.427..728D,2013MNRAS.432..886D,2014MNRAS.439.2970D}). 

        On the other hand, asteroid 2019~AQ$_{3}$ is not currently in mean-motion resonance with any planet; it is, however, in 
        near-mean-motion resonance with Mercury (15:8), Venus (11:15), the Earth (9:20), Mars (6:25), and Jupiter (1:26) ---the ones with 
        Mars, Mercury, and Jupiter (in this order) are the closest. In Section~\ref{source}, we show that the Earth--Moon system and Jupiter 
        are the main secular perturbers of 2019~AQ$_{3}$.

\begin{figure}
       \centering
        \includegraphics[width=\linewidth]{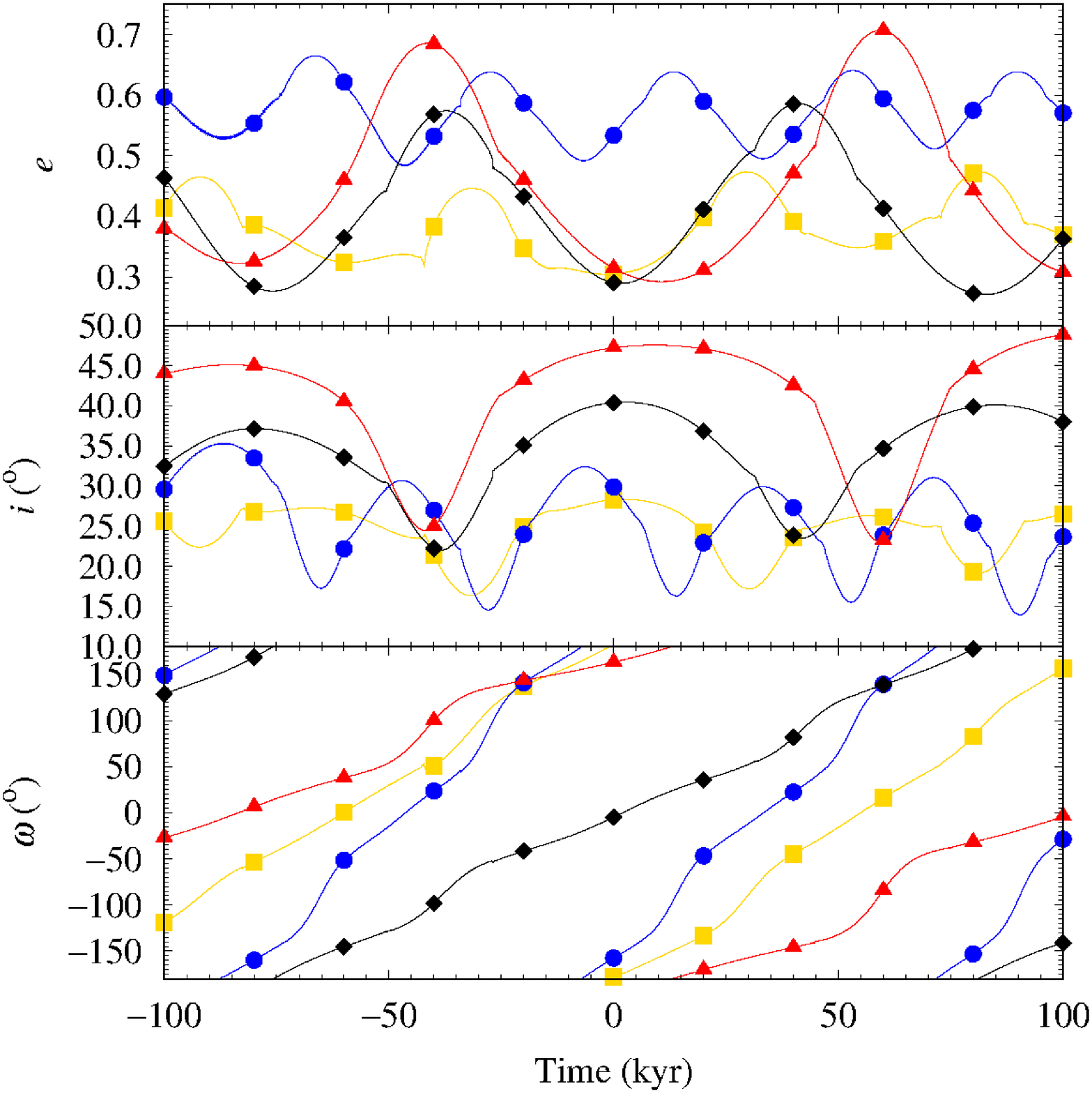}
        \caption{Evolution of the values of the eccentricity (top panel), inclination (middle panel), and argument of perihelion (bottom 
                 panel) of the nominal orbits of representative Atira-class asteroids: (418265) 2008~EA$_{32}$ (gold squares), 
                 2010~XB$_{11}$ (blue circles), 2018~JB$_{3}$ (black diamonds), and 2019~AQ$_{3}$ (red triangles).
                }
        \label{atiras}
     \end{figure}

        Figure~\ref{atiras} shows the short-term evolution of $e$, $i$, and $\omega$ of 2019~AQ$_{3}$ (red triangles) within the context of 
        those of other Atiras ---(418265) 2008~EA$_{32}$ (gold squares), 2010~XB$_{11}$ (blue circles), and 2018~JB$_{3}$ (black diamonds)---  
        also displaying the behaviour pointed out above (see also fig.~1 in \citealt{2018RNAAS...2b..46D}). It cannot be discarded that a 
        sizeable population of stable (in the sense of remaining confined within a relatively small section of the orbital parameter space) 
        Atiras/Vatiras may share high values of the inclination, particularly if their arguments of perihelia are close to 90\degr or 
        270\degr so they avoid close encounters with Venus at aphelion. 
%
%
%
%

\begin{figure*}
       \centering
        \includegraphics[width=0.49\linewidth]{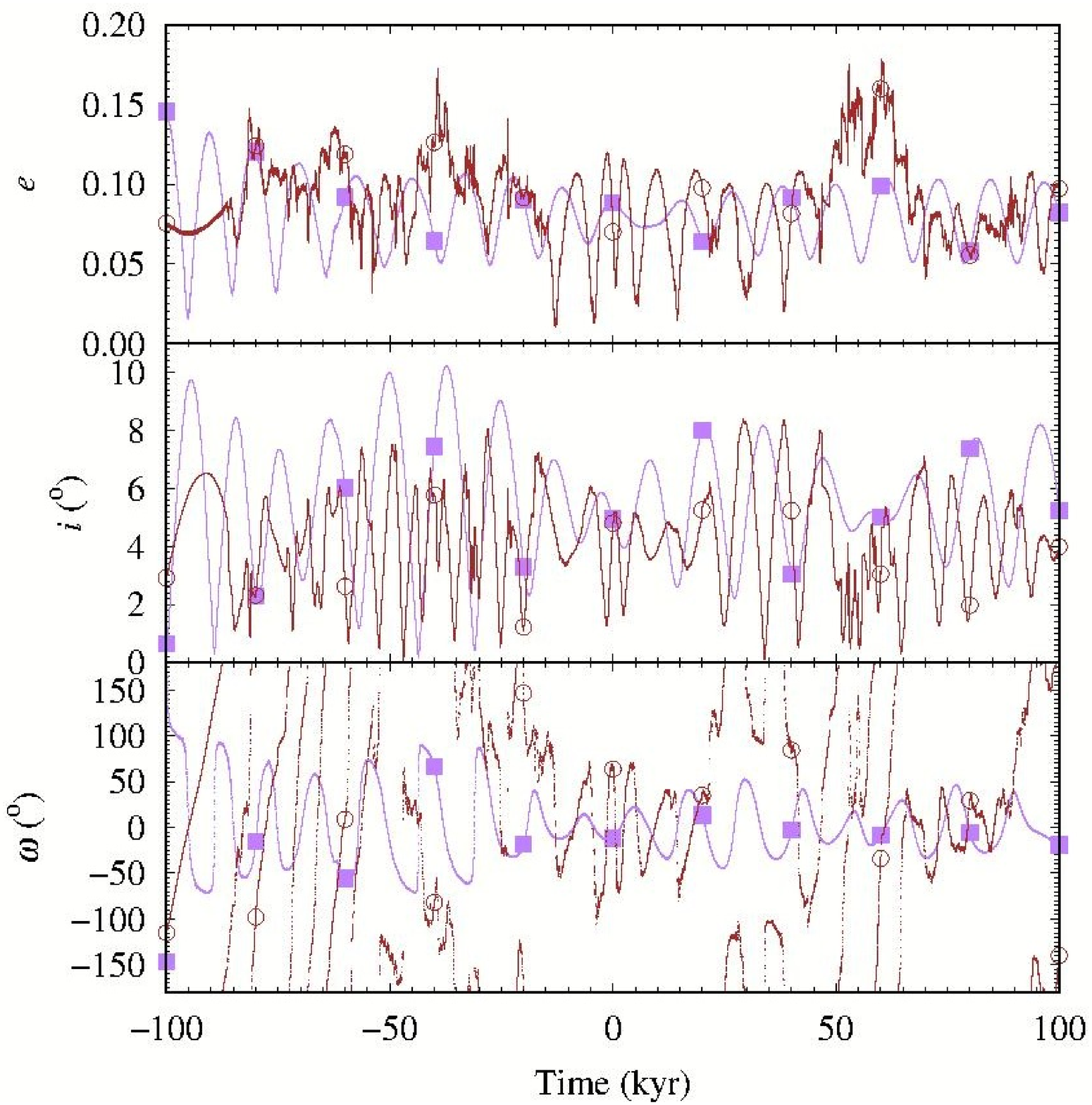}
        \includegraphics[width=0.49\linewidth]{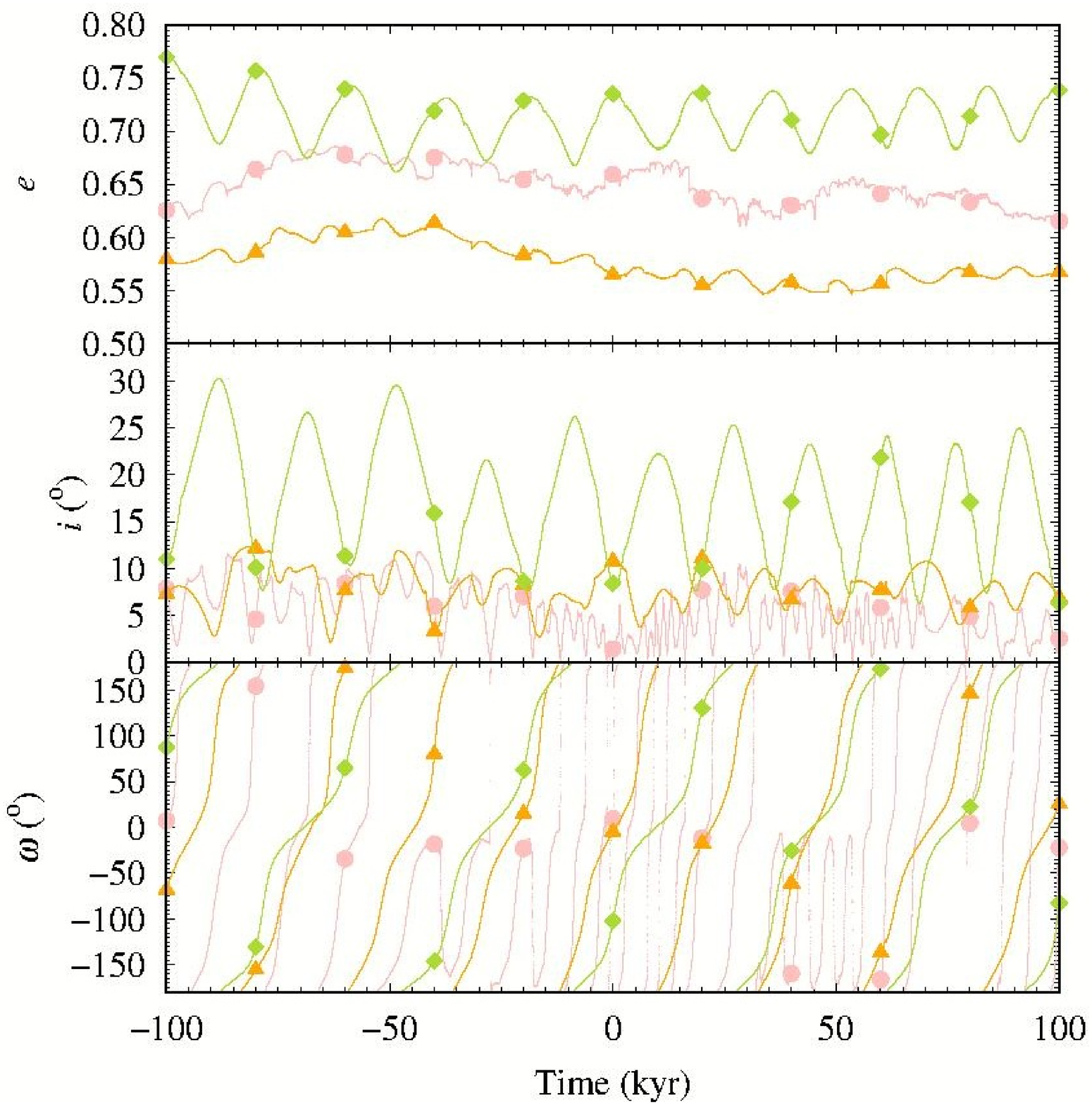}
        \caption{As Fig.~\ref{atiras} but for the nominal orbits of selected Atira, Aten, and Apollo asteroids: 2012~FC$_{71}$ (purple 
                 squares, Aten, left-hand side panels), 2013~JX$_{28}$ (orange triangles, Atira, right-hand side panels), (459872) 
                 2014~EK$_{24}$ (brown empty circles, Apollo, left-hand side panels), 2016~XK$_{24}$ (green diamonds, Aten, right-hand side 
                 panels), and 2019~BE$_{5}$ (pink filled circles, Aten, right-hand side panels).
                }
        \label{LK}
     \end{figure*}

        If the known Atiras are not subjected to a present-day Lidov--Kozai resonance, they should exhibit obvious differences when compared 
        with other NEAs truly engaged in Lidov--Kozai resonant behaviour. Figure~\ref{LK} shows the evolution of the relevant parameters of 
        representative minor bodies. NEAs 2012~FC$_{71}$ (purple squares in Fig.~\ref{LK}, left-hand side panels) and (459872) 2014~EK$_{24}$ 
        (brown empty circles in Fig.~\ref{LK}, left-hand side panels) are clear examples of Lidov--Kozai evolution, particularly in the case 
        of Aten-class asteroid 2012~FC$_{71}$. The dynamical evolution of Apollo-class ($a>1$~au and $q<1.017$~au) asteroid 459872 is far 
        more chaotic (the values of the various parameters change more rapidly and by a wider margin), but still displaying clear episodes 
        of Lidov--Kozai resonant behaviour. In both cases, the value of the orbital inclination is low and the argument of perihelion tends 
        to librate around 0\degr or 180\degr. The evolution of both objects has been studied in detail by \citet{2015A&A...580A.109D}. 

        The evolution of Atira-class asteroid 2013~JX$_{28}$ (orange triangles in Fig.~\ref{LK}, right-hand side panels) and Aten-class 
        asteroid 2016~XK$_{24}$ (green diamonds in Fig.~\ref{LK}, right-hand side panels) is somewhat similar to those presented in 
        Fig.~\ref{atiras}, although the inclinations are lower and the eccentricities higher. Asteroid 2016~XK$_{24}$ has the lowest value 
        of the sidereal orbital period and 2013~JX$_{28}$ the third lowest. Aten-class asteroid 2016~XK$_{24}$ shows that the widespread 
        dynamical behaviour found for the Atiras ---i.e. coupled oscillation of the values of eccentricity and inclination, but no libration 
        of the value of the argument of perihelion--- is not exclusive of this dynamical class. Rather puzzling is the evolution of the 
        recently discovered Aten-class asteroid 2019~BE$_{5}$ \citep{2019MPEC....C...10O},\footnote{\url{https://minorplanetcenter.net/mpec/K19/K19C10.html}} 
        which exhibits libration of the argument of perihelion around $-$90\degr (or 270\degr) and has the fourth lowest value of the 
        semimajor axis among known bound minor bodies, 0.61019$\pm$0.00011~au (pink filled circles in Fig.~\ref{LK}, right-hand side panels), 
        and a minimum orbit intersection distance, or MOID, with the Earth of 0.000163~au. The Lidov--Kozai resonant behaviour displayed by 
        2019~BE$_{5}$ is puzzling because it has a low value of the orbital inclination during the resonant episodes; the object follows a 
        rather eccentric orbit and this fact may explain why $\omega$ does not librate around 0\degr or 180\degr. Its orbit determination is 
        rather uncertain though as it is based on 76 observations with a data-arc span of 8~d.   
%
%
%
%

     \subsection{Short-term orbital evolution with respect to the invariable plane}\label{invar} 
\begin{figure}
       \centering
        \includegraphics[width=\linewidth]{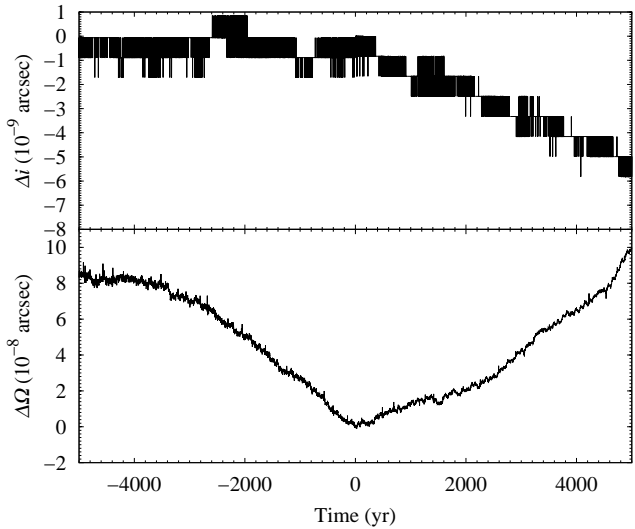}
        \caption{Temporal variations $\Delta{i}$ (top panel) and $\Delta\Omega$ (bottom panel) in the orientation of the invariable plane of 
                 the system. $\Delta{i}=0''$ ($\Delta\Omega=0''$) corresponds to $i=1\fdg57809813$ ($\Omega=107\fdg58872717$) at $t=0$.  
                }
        \label{inv}
     \end{figure}

\begin{figure}
       \centering
        \includegraphics[width=\linewidth]{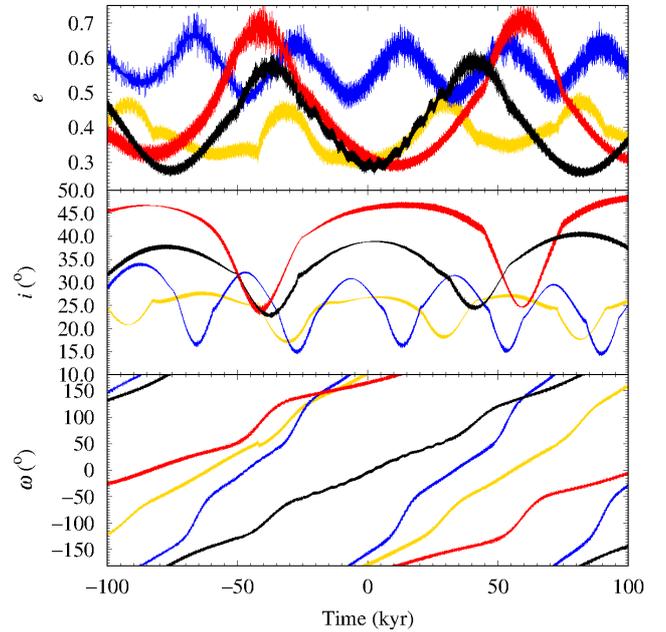}
        \caption{As Fig.~\ref{atiras} but with orbital elements computed in the frame of reference of the invariable plane of the system. 
                }
        \label{atirasINV}
     \end{figure}

        So far, we have computed the orbital elements with respect to the ecliptic which is particular to the Earth and not significant for 
        the entire Solar system. A dynamically more relevant plane is in the invariable plane of the Solar system that is defined as the 
        plane through the barycentre that is normal to the angular momentum vector of the Solar system, which is a constant of the motion. 
        \citet{2012A&A...543A.133S} have shown that the ecliptic is within a few degrees of the invariable plane; the situation is 
        equivalent to the one depicted in their fig.~1. The transformation between coordinates and velocities in the ecliptic frame of 
        reference to the invariable plane is carried out by rotating first an angle $\Omega_{\rm inv}$ about the $z$-axis and then an angle
        $i_{\rm inv}$ about the $x$-axis ---rotation matrices in \citet{2012A&A...543A.133S} replacing $\varphi$ with $\Omega_{\rm inv}$ and
        $\epsilon$ with $i_{\rm inv}$. The osculating values of $\Omega_{\rm inv}$ and $i_{\rm inv}$ have been obtained using the equations
        (5) and (6) in \citet{2012A&A...543A.133S} that make use of equations (1) to (4) in section 2.2 of the same work. At output time, we 
        have computed the instantaneous values of $\Omega_{\rm inv}$ and $i_{\rm inv}$ using these equations, and changed the position and 
        velocity vectors to the invariable plane before computing the orbital elements in the new frame of reference. At $t=0$, 
        $i_{\rm inv}=1\fdg57809813$ and $\Omega_{\rm inv}=107\fdg58872717$; the values obtained by \citet{2012A&A...543A.133S} are 
        $i_{\rm inv}=1{\degr}34'43\farcs3$ (1\fdg57870312) and $\Omega_{\rm inv}=107{\degr}34'56''$ (107\fdg58227198) at the epoch J2000.0. 
        Figure~\ref{inv} shows the short-term accuracy of our determination of the orientation of the invariable plane; therefore, our 
        calculations are accurate enough to obtain reliable conclusions regarding the present-day operation (or not) of the Lidov--Kozai 
        resonance in the case of known Atiras. Figure~~\ref{atirasINV} shows the evolution of the values of $e$, $i$, and $\omega$ in the 
        invariable reference frame and no libration of $\omega$ is observed. This result confirms that there are no known Atiras currently 
        trapped in a Lidov--Kozai resonance, but Figs~\ref{evolution} and \ref{control} clearly show that this situation may change in the 
        future.
%
%
%
%
%
%
%
%

     \subsection{Medium-term evolution}
        Figure~\ref{evolution}, right-hand side set of panels, shows the medium-term evolution (from 10~Myr into the past to 10~Myr into the 
        future) of the values of relevant parameters for the nominal orbit of 2019~AQ$_{3}$. Close encounters with both Venus and Mercury 
        are systematically avoided after about 3.5~Myr into the future (but also prior to 8~Myr into the past) and the value of the 
        Lidov--Kozai parameter changes within a smaller interval as well. On this extended time-scale, the value of the semimajor axis 
        (fourth to top panel) changes by about 20 per cent (i.e. the overall behaviour is obviously chaotic), although after about 3.5~Myr 
        it remains nearly constant. Now Mercury becomes the main direct perturber, but the Earth--Moon system and Jupiter remain as distant
        secular perturbers. 

        \subsubsection{Lidov--Kozai evolution}\label{LKev}
           After about 3.5~Myr into the future, the dynamical context changes significantly; now Mercury becomes the main direct perturber. 
           Prior to this, a close encounter with Venus leads 2019~AQ$_{3}$ to a rather smooth path with nearly constant semimajor axis that 
           is no longer within the Atira orbital realm but within the Vatira one (in thick red/grey, fourth to top panel). Now the coupled 
           oscillation of the three orbital parameters, $e$, $i$, and $\omega$, is observed. The value of the argument of perihelion mostly 
           librates about 270\degr. In other words, the aphelion takes place when 2019~AQ$_{3}$ is the farthest from the ecliptic plane or 
           aphelion always occurs away from the orbital plane of Venus. The Lidov--Kozai resonance is in effect, protecting 2019~AQ$_{3}$ 
           from close encounters with Venus. 
%
%
     \begin{figure*}
       \centering
        \includegraphics[width=0.49\linewidth]{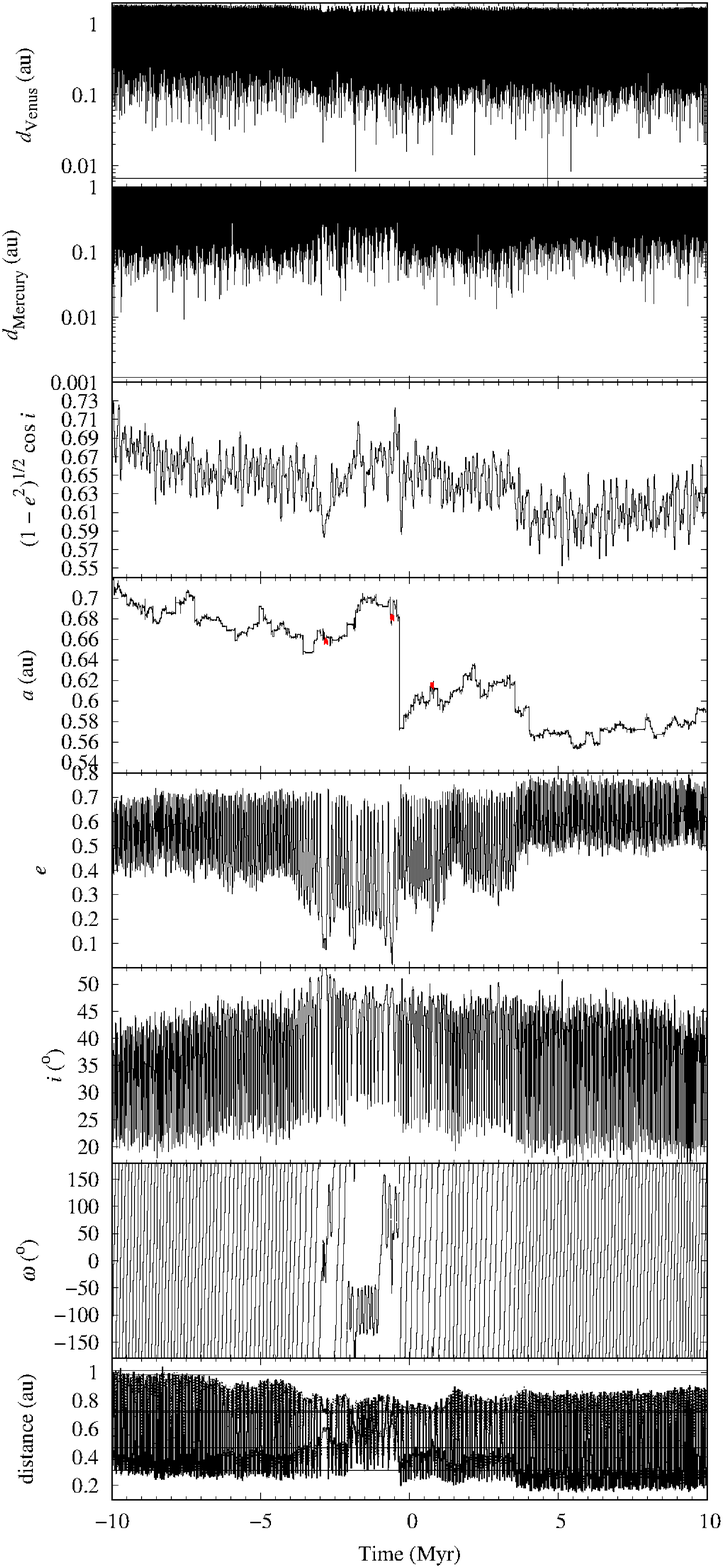}
        \includegraphics[width=0.49\linewidth]{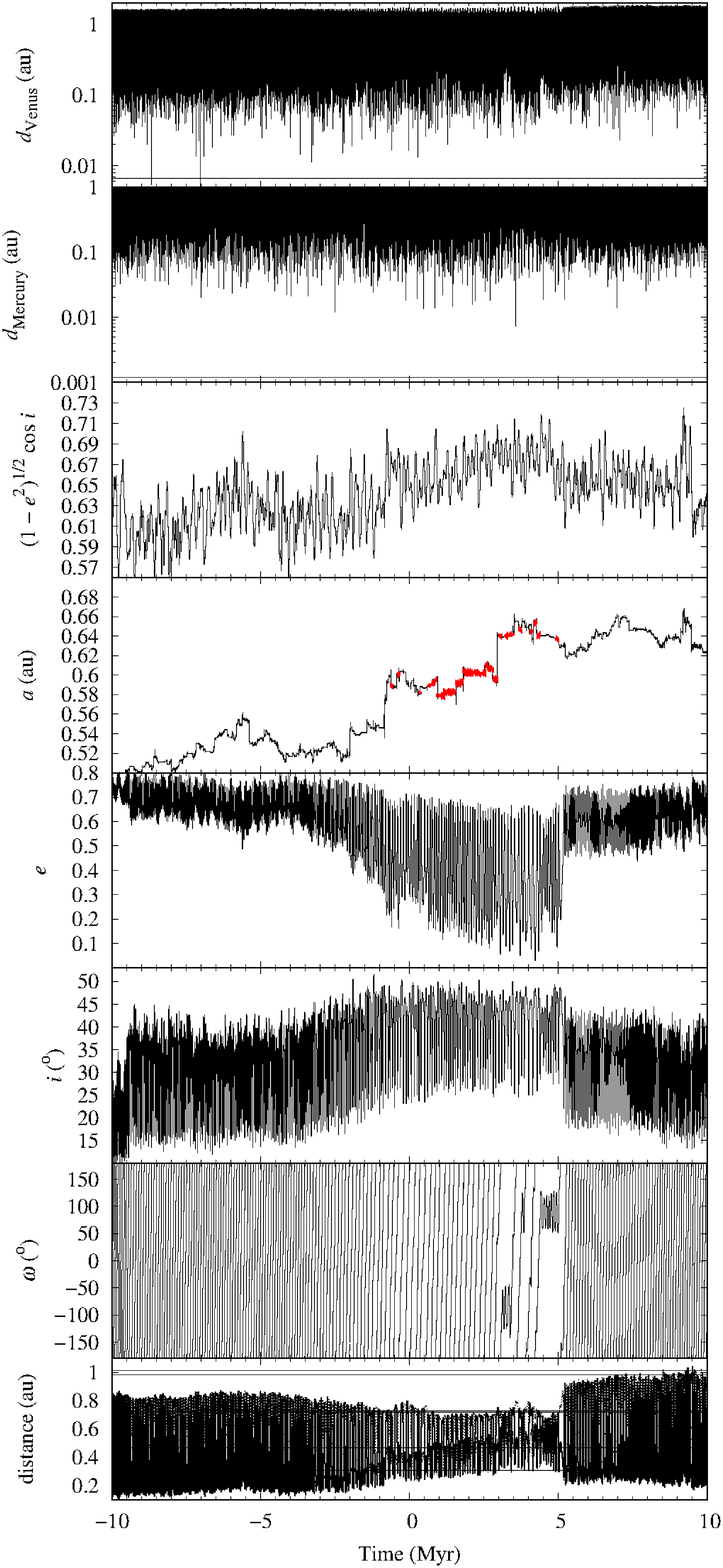}
        \caption{As Fig.~\ref{evolution} but for two control orbits that are most different from the nominal one (yet marginally compatible 
                 with the observations, see the text for details).
                }
        \label{control}
     \end{figure*}
%
%

           Figure~\ref{control} is analogous to Fig.~\ref{evolution} but shows the long-term evolution of two relevant control orbits based 
           on the nominal solution but adding (+, right-hand set) or subtracting ($-$, left-hand set) three times the corresponding 
           uncertainty from each orbital element (the six of them) in Table~\ref{elements}. These two examples of orbit evolution using 
           initial conditions that are very different from those associated with the nominal orbit (yet still marginally compatible with the 
           observations) are not meant to show how large the dispersion of the various parameters could be as they change over time, but to 
           indicate that the onset of the Lidov--Kozai resonance is not only restricted to the nominal orbit or its immediate dynamical 
           neighbourhood. Episodes of Lidov--Kozai resonance lasting for more than 1~Myr are observed in both cases.

           \citet{2016ARA&A..54..441N} has studied the Lidov--Kozai mechanism and how it can secularly excite the value of the eccentricity 
           until collisions with inner or outer perturbers are possible. Although the orbital architecture driving the evolution of 
           2019~AQ$_{3}$ appears to contribute to enhance the stability of the orbits in the sense of keeping them within a well-defined
           section of the available orbital parameter space, Fig.~\ref{evolution} indicates that the eccentricity may eventually reach 
           values that lead the aphelion closer to the Earth--Moon system, perhaps triggering an impact or an ejection from NEA parameter 
           space. Conversely, a larger value of the eccentricity may lead to a shorter perihelion and an eventual collision with the Sun or 
           an ejection from the Solar system or into its outskirts. In our case, the value of the aphelion distance, that oscillates between 
           those of the semimajor axes of Venus and the Earth, can become as high as 1.02~au (see Fig.~\ref{control}, right-hand panel), 
           opening the door to eventual impacts on our planet if one of the nodes reaches aphelion (see also \citealt{2016MNRAS.458.4471R}). 
           Members of the Atira population can eventually become impactors as a result of the complex orbital evolution discussed here and 
           in the event of an impact, they will most probably come from out of the Sun's glare. 

        \subsubsection{Vatira evolution}
           The evolution of 2019~AQ$_{3}$ shown in Fig.~\ref{evolution} initially leads to brief periods (in thick red/grey) in which the
           aphelion distance of 2019~AQ$_{3}$, $Q<0.718$~au i.e. 2019~AQ$_{3}$, an Atira, becomes a transient Vatira. This opens the door to
           transitions between the two dynamical classes. In fact, the dynamical context that keeps 2019~AQ$_{3}$ in a chaotic but 
           relatively stable state may be effective in enabling a dynamical pathway to go from the Atira orbital realm into the Vatira one 
           as shown by the evolution after about 3.5~Myr into the future. The progressive reduction in the value of the semimajor axis 
           eventually leads to a transition from the Atira population into the Vatiras that avoid encounters with Venus altogether. 
           Figure~\ref{control} shows that transitions to and from the Vatira orbital realm are also possible well away from the nominal 
           orbit. Our calculations show that a combination of secular evolution and eventual close encounters with Venus triggers the
           transitions.

           Atiras do not cross the orbit of the Earth--Moon system and Vatiras do not cross the orbit of Venus; both Atiras and Vatiras 
           define somewhat disjoint subsets. Atiras can impact on Venus, Vatiras cannot; therefore and although the members of both 
           dynamical classes interact with Venus, they do it quite differently. We may now argue that the distinctive dynamical property of 
           putative, long-term stable Vatiras is that they must be virtually and dynamically detached from Venus. This implies a robust 
           difference between the dynamical behaviour of hypothetically stable Atiras, such as those discussed by 
           \citet{2016MNRAS.458.4471R}, and stable Vatiras. In general, Atiras may still experience close encounters with Venus, Vatiras 
           cannot. When the Lidov--Kozai resonance is at work, in addition to avoiding close encounters with Venus at aphelion, close 
           encounters with Mercury at perihelion are also reduced, but the orbit of Mercury is far more eccentric (0.21) and inclined 
           (7\fdg0) than that of Venus. This aspect is more obvious when the evolution of the nodal distances is studied (bottom panels in 
           Figs~\ref{evolution} and \ref{control}). After about 3.5~Myr into the future, the orbital nodes remain well away from the path of 
           Venus, which must improve the overall orbital stability of 2019~AQ$_{3}$. We observe that this situation may have happened in the 
           past, but a close encounter with Venus nearly 5.5~Myr ago led to an extended period of time in which the nodes of 2019~AQ$_{3}$ 
           crossed regularly the path of Venus; however, 2019~AQ$_{3}$ was not a Vatira at that time (see Fig.~\ref{evolution}, right-hand 
           side panels).

     \subsection{Dominant perturbers of the short-term dynamics}\label{source}
        The textbook description of the Lidov--Kozai mechanism (see e.g. \citealt{1999ssd..book.....M}) includes the coupled oscillation of 
        the values of $e$ and $i$, in such a way that when the eccentricity is at its highest, the inclination is at its lowest (and vice 
        versa), and the argument of perihelion librates about 0{\degr} or 180{\degr} when the value of the inclination is low and about 
        90{\degr} or 270{\degr} when the inclination is high; this keeps the value of the Lidov--Kozai parameter (see above) nearly 
        constant. However, for 2019~AQ$_{3}$, Fig.~\ref{evolution}, left-hand side set of panels, suggests a more complicated present-day 
        arrangement that is still able to produce the familiar oscillation in $e$ and $i$, but not in $\omega$.

\begin{figure}
       \centering
        \includegraphics[width=\linewidth]{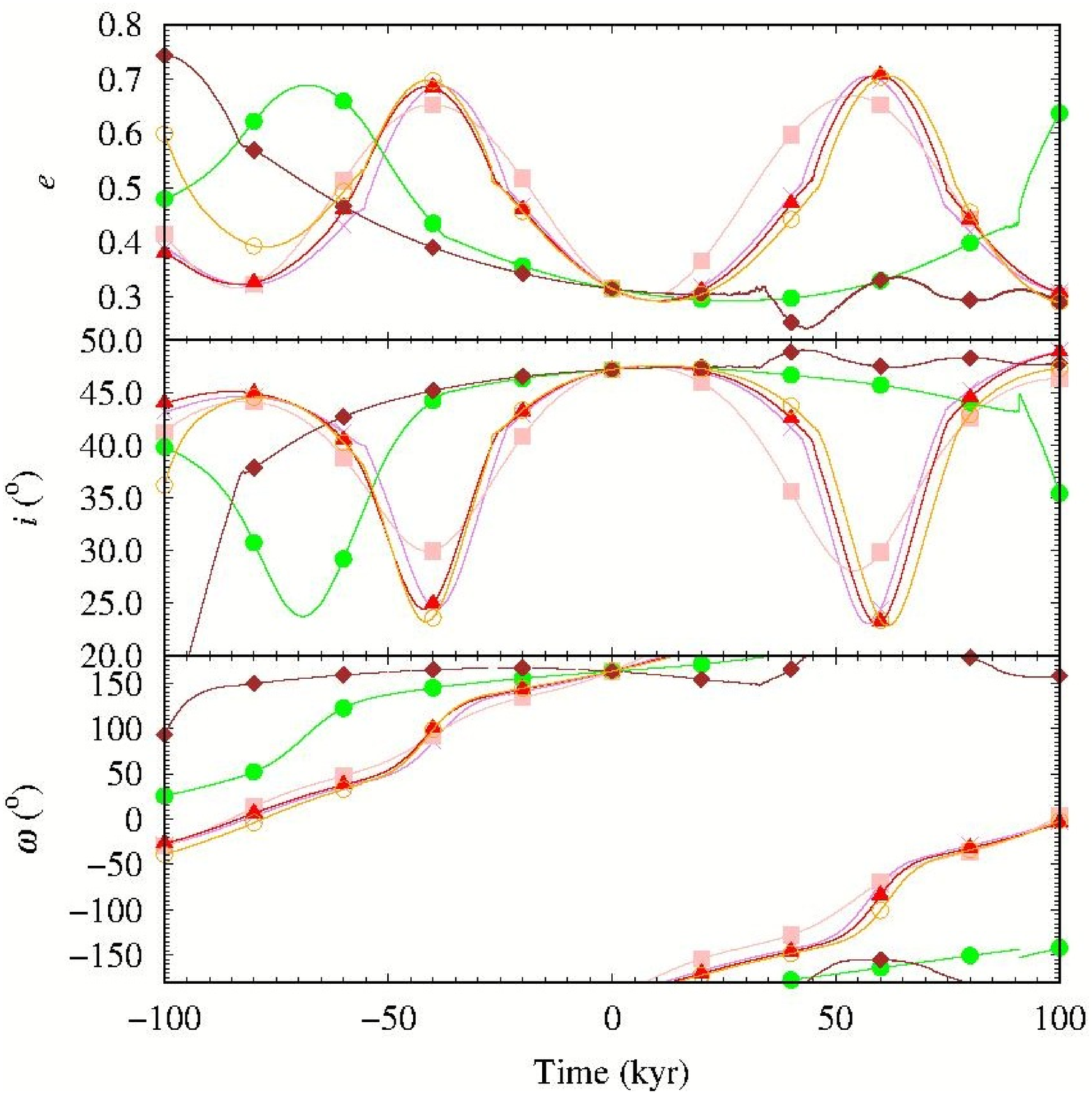}
        \caption{Evolution of the values of the eccentricity (top panel), inclination (middle panel), and argument of perihelion (bottom 
                 panel) of the nominal orbit of 2019~AQ$_{3}$ under the usual physical model (red triangles), without Mercury (violet 
                 $\times$), without Venus (pink squares), without the Earth--Moon system (green filled circles), without Jupiter (brown 
                 diamonds), and without Saturn (orange empty circles).
                }
        \label{kozai}
     \end{figure}

\begin{figure}
       \centering
        \includegraphics[width=\linewidth]{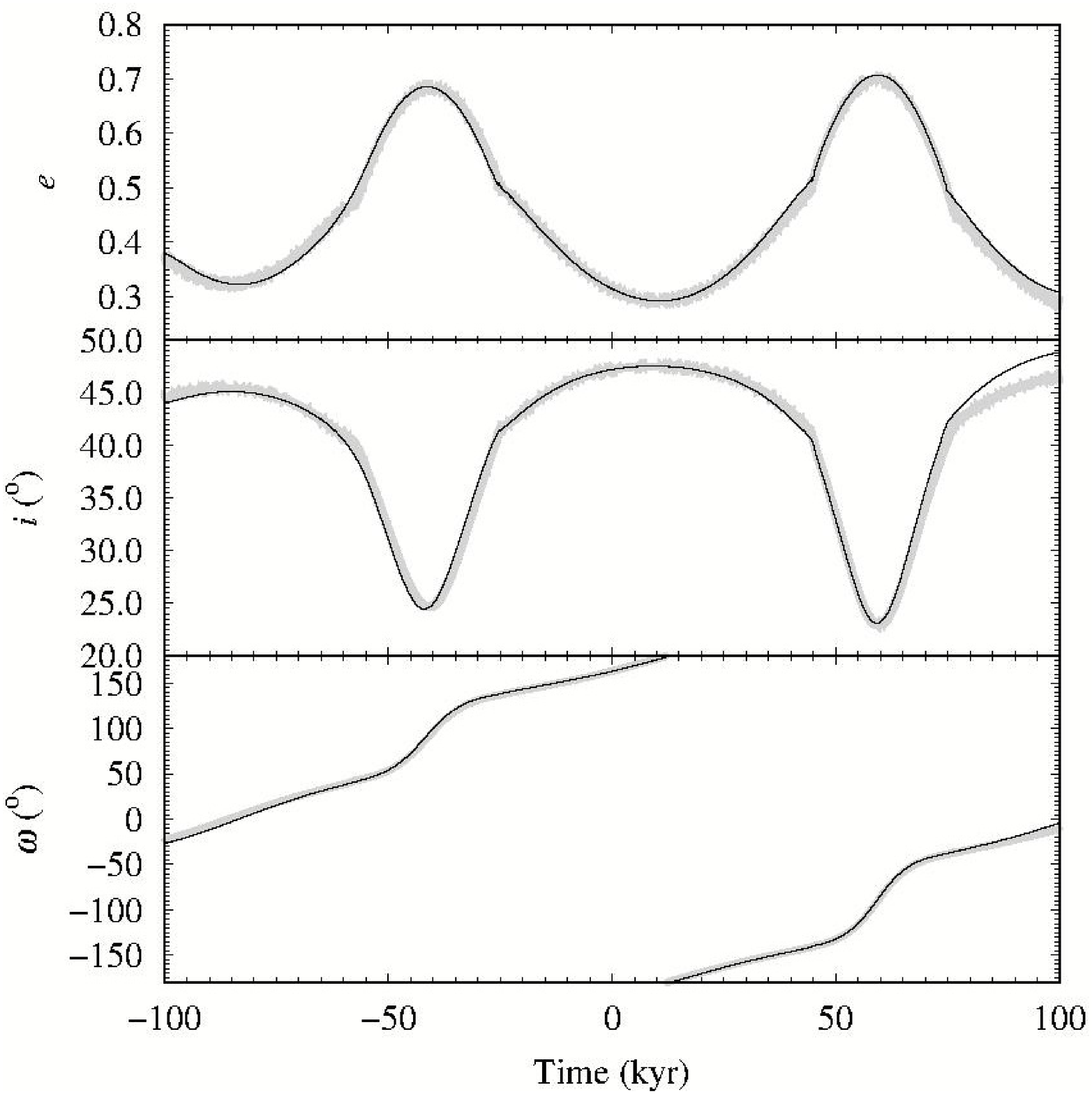}
        \caption{As Fig.~\ref{kozai}, nominal orbit of 2019~AQ$_{3}$ under the usual physical model (solid black) and only with Venus, the 
                 Earth--Moon system, and Jupiter (thick solid light grey).  
                }
        \label{3perts}
     \end{figure}

        In our case, 2019~AQ$_{3}$ might be controlled by two direct perturbers, Venus and the Earth--Moon system, and a distant one, 
        Jupiter. In order to test which perturber is actually responsible for the observed orbital evolution, we have performed additional 
        simulations without the various planets. Relevant results from these simulations are shown in Fig.~\ref{kozai}. When Venus is 
        removed, the orbital evolution of 2019~AQ$_{3}$ changes only slightly and becomes smooth (pink squares curve in Fig.~\ref{kozai}) as 
        the kinks disappear (compare pink squares and red triangles curves). We interpret this result as strong evidence in favour of Venus 
        being a secondary actor in this case. However, when either the Earth--Moon system (green filled circles curve in Fig.~\ref{kozai}) 
        or Jupiter (brown diamonds curve in Fig.~\ref{kozai}) are removed from the calculations, the impact on the orbital evolution of 
        2019~AQ$_{3}$ is such that we can consider the coupled oscillatory behaviour in $e$ and $i$ as effectively broken. When either 
        Mercury (violet $\times$ curve in Fig.~\ref{kozai}) or Saturn (orange empty circles curve in Fig.~\ref{kozai}) are removed from the 
        calculations, no significant effects are seen (i.e. their contribution is weaker than that of Venus). Figure~\ref{3perts} shows the 
        evolution when only the Sun, Venus, the Earth--Moon system, and Jupiter are included in the calculations. The simplified system of 
        three perturbers reproduces the observed evolution better than other arrangements and therefore we consider our interpretation as 
        confirmed: the observed dynamical evolution is the result of the combined action of two dominant perturbers, the Earth--Moon system 
        and Jupiter, and a secondary one, Venus. 

%
%
%
%
%
%
%
%

     \subsection{Post-Newtonian evolution}\label{relcal}
\begin{figure}
       \centering
        \includegraphics[width=\linewidth]{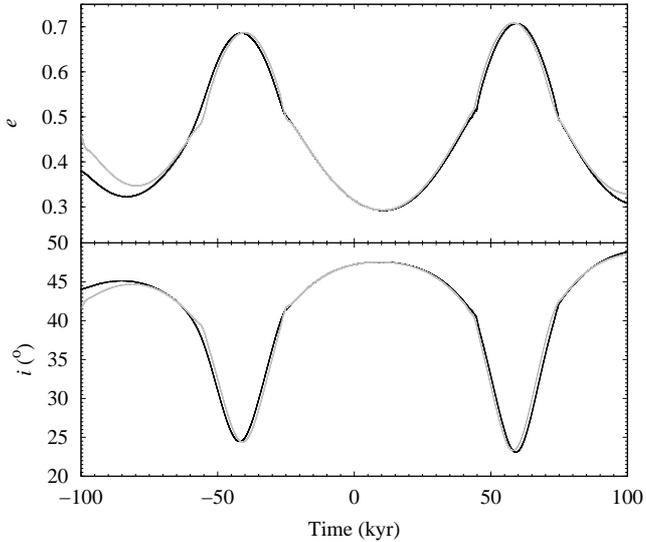}
        \caption{Evolution of the values of the eccentricity (top panel) and inclination (bottom panel) of the nominal orbit of 
                 2019~AQ$_{3}$ under the Newtonian (in black) and post-Newtonian (in grey) approximations.
                }
        \label{PN}
     \end{figure}

        Because of the effect of the secular dynamics discussed above, the perihelion distance of 2019~AQ$_{3}$ is always in the 
        neighbourhood or inside the orbit of Mercury and can become as low as 0.17~au, below that of asteroids such as 1566~Icarus (1949~MA) 
        for which relativistic motion in the form of non-Newtonian perihelion precession has been measured (see e.g. 
        \citealt{1992AJ....104.1226S}). However, the overall results of our calculations under the post-Newtonian approximation are 
        consistent with those obtained in the Newtonian case (see Fig.~\ref{PN}). The post-Newtonian evolution (in grey) preserves the 
        coupled oscillations in $e$ and $i$ observed under the Newtonian approximation (in black). The issue of general relativistic 
        precession operating together with Lidov--Kozai oscillations has been explored by \citet{2017MNRAS.468.1405S}. Although 2019~AQ$_{3}$
        is not currently trapped in a Lidov--Kozai resonance, the overall evolution is preserved in the post-Newtonian case. 
%
%
%
%

  \section{Discussion}
\begin{figure}
       \centering
        \includegraphics[width=\linewidth]{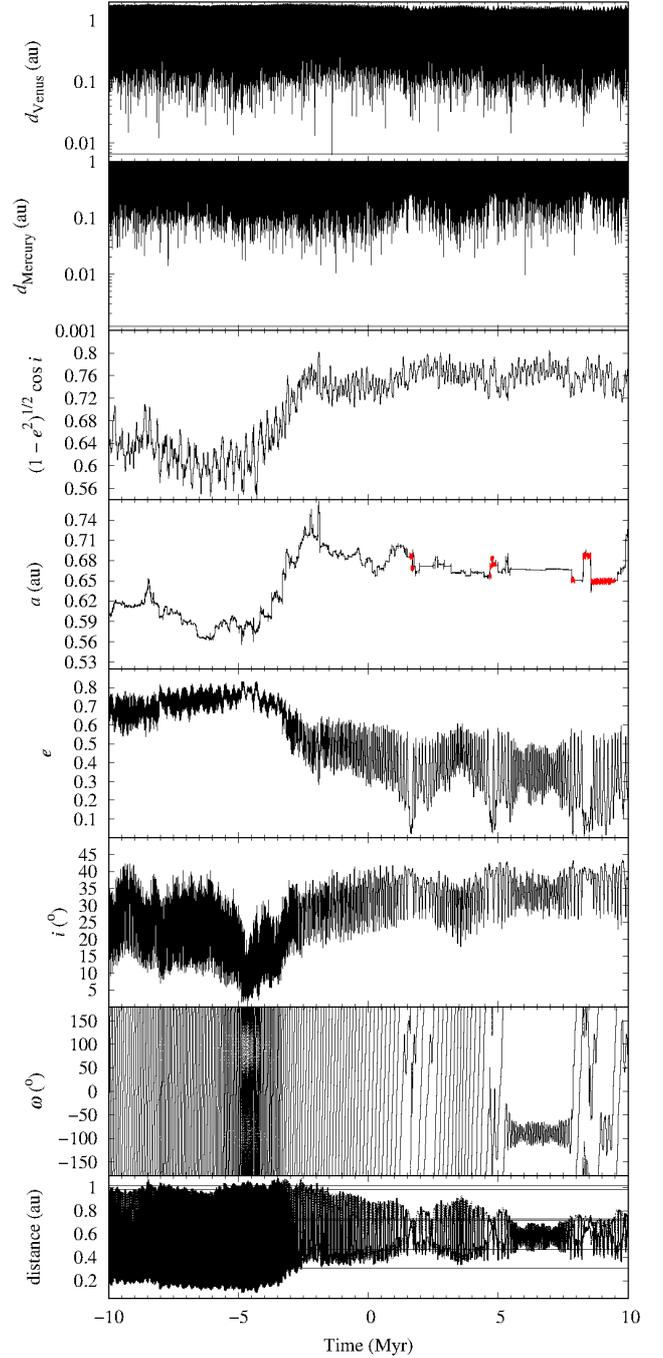}
        \caption{As Fig.~\ref{evolution} but for the nominal orbit of 2018~JB$_{3}$.
                }
        \label{2018jb3}
     \end{figure}

     The data in Section 2.1 hint at a possible outlier nature of 2019~AQ$_{3}$ both within the known NEAs and even among the known Atiras.
     Table~\ref{knownatiras} shows the available orbital data for NEAs currently classified as members of the Atira dynamical class. On the 
     one hand and considering the statistical parameters Q$_{3}$, third quartile, IQR, interquartile range, and OU, upper outlier limit 
     (Q$_{3}+1.5$ IQR, \citealt{1977eda..book.....T}), for the orbital inclination of the 19 known Atiras we obtain: Q$_{3}$=26\fdg94, 
     IQR=13\fdg36, and OU=46\fdg99. With a value of the inclination of 47\fdg2186$\pm$0\fdg0009, 2019~AQ$_{3}$ barely qualifies as an 
     outlier; however, it certainly is a strong outlier in terms of aphelion distance (0.773684~au versus OL=0.825768~au) ---(418265) 
     2008~EA$_{32}$ is another outlier in $Q$ (0.803780~au). On the other hand, a statistical analysis analogous to the one presented by 
     \citet{2019MNRAS.483L..37D} that used the orbit model\footnote{\url{http://neo.ssa.esa.int/neo-population}} developed by the 
     Near-Earth Object Population Observation Program (NEOPOP) and described by \citet{2018Icar..312..181G} shows that 2019~AQ$_{3}$ is 
     indeed a statistical outlier within the known NEA population. Because of its high orbital inclination, there are no other known 
     objects with similar orbit determinations. The orbit model predicts the existence of five objects (for the entire NEO population with 
     $H<25$~mag, about 800\,000 predicted NEOs) orbitally similar to 2019~AQ$_{3}$ but with absolute magnitudes, $H$, in the range 
     23--25~mag, considerably smaller than 2019~AQ$_{3}$. Therefore, it is an outlier within the known NEO orbits (perhaps due to 
     observational bias, see e.g. \citealt{2016MNRAS.458.4471R}), but also an outlier (in terms of size) when considering the predictions 
     from a state-of-the-art orbit model. Regarding the overall size distribution of the known Atiras, there is an unusually large number of 
     objects with $H<$23--25~mag (see last column in Table~\ref{knownatiras}). Figure~\ref{2018jb3} shows the evolution of 2018~JB$_{3}$ 
     that goes from Aten to Atira and eventually becomes a Vatira. This dynamical evolution suggests that the Atira/Vatira orbital realm may 
     be repopulated with former Atens.
%
%
      \begin{table*}
        \centering
        \fontsize{8}{11pt}\selectfont
        \tabcolsep 0.20truecm
        \caption{Heliocentric orbital elements and parameters ---$q=a(1-e)$ is the perihelion distance, $Q=a(1+e)$ is the aphelion 
                 distance--- of the known Atiras. The statistical parameters are Q$_{1}$, first quartile, Q$_{3}$, third quartile, IQR, 
                 interquartile range, OL, lower outlier limit (Q$_{1}-1.5\ $IQR), and OU, upper outlier limit (Q$_{3}+1.5\ $IQR). The orbit 
                 determinations are referred to epoch JD 2458600.5 (2019-Apr-27.0) TDB (J2000.0 ecliptic and equinox) with the exception of
                 1998~DK$_{36}$ (JD 2450868.5 TDB) and 2015~ME$_{131}$ (JD 2457197.5 TDB). Source: JPL's SBDB.
                }
        \begin{tabular}{lrrrrrrrr}
          \hline
             Object             & $a$ (au) & $e$      & $i$ (\degr) & $\Omega$ (\degr) & $\omega$ (\degr) & $q$ (au) & $Q$ (au) & $H$ (mag) \\
          \hline
                 (163693) Atira & 0.740919 & 0.322122 & 25.620193   & 103.888396       & 252.886331       & 0.502253 & 0.979585 & 16.3      \\
       (164294) 2004~XZ$_{130}$ & 0.617615 & 0.454551 &  2.950756   & 211.373871       &   5.180663       & 0.336878 & 0.898353 & 20.4      \\
        (413563) 2005~TG$_{45}$ & 0.681393 & 0.372181 & 23.336466   & 273.436802       & 230.416868       & 0.427792 & 0.934994 & 17.6      \\
        (418265) 2008~EA$_{32}$ & 0.615907 & 0.305035 & 28.264395   & 100.958641       & 181.840605       & 0.428034 & 0.803780 & 16.4      \\
         (434326) 2004~JG$_{6}$ & 0.635232 & 0.531142 & 18.943867   &  37.029381       & 352.998487       & 0.297834 & 0.972631 & 18.4      \\
        (481817) 2008~UL$_{90}$ & 0.695049 & 0.379850 & 24.309761   &  81.142712       & 183.642522       & 0.431034 & 0.959063 & 18.6      \\
                 1998~DK$_{36}$ & 0.692257 & 0.416017 &  2.017520   & 151.461720       & 180.042707       & 0.404267 & 0.980248 & 25.0      \\
                  2006~WE$_{4}$ & 0.784745 & 0.182918 & 24.767078   & 311.014633       & 318.597466       & 0.641201 & 0.928288 & 18.9      \\
                 2010~XB$_{11}$ & 0.618029 & 0.533873 & 29.886574   &  96.315122       & 202.484306       & 0.288080 & 0.947978 & 19.9      \\
                 2012~VE$_{46}$ & 0.713043 & 0.361305 &  6.675782   &   8.784572       & 190.539803       & 0.455417 & 0.970669 & 20.2      \\
                 2013~JX$_{28}$ & 0.600855 & 0.564131 & 10.761830   &  39.955081       & 354.887304       & 0.261894 & 0.939815 & 20.1      \\
                  2013~TQ$_{5}$ & 0.773707 & 0.155579 & 16.398688   & 286.770884       & 247.296796       & 0.653334 & 0.894079 & 19.8      \\
                 2014~FO$_{47}$ & 0.752123 & 0.271100 & 19.197106   & 358.653061       & 347.463144       & 0.548222 & 0.956023 & 20.3      \\
                2015~DR$_{215}$ & 0.666400 & 0.471499 &  4.088461   & 314.953867       &  42.283540       & 0.352193 & 0.980607 & 20.3      \\
                2015~ME$_{131}$ & 0.806118 & 0.204225 & 30.236744   & 315.351877       & 162.273931       & 0.641489 & 0.970747 & 19.5      \\
                  2017~XA$_{1}$ & 0.809618 & 0.201567 & 17.176997   & 239.659585       & 327.622064       & 0.646426 & 0.972811 & 21.2      \\
                        2017~YH & 0.634446 & 0.482474 & 19.833554   & 134.222643       & 147.454605       & 0.328342 & 0.940550 & 18.5      \\
                  2018~JB$_{3}$ & 0.683189 & 0.290494 & 40.392589   & 106.425908       & 355.237492       & 0.484726 & 0.881651 & 17.6      \\
                  2019~AQ$_{3}$ & 0.588662 & 0.314310 & 47.218559   &  64.487327       & 163.151818       & 0.403639 & 0.773684 & 17.6      \\
          \hline
                           Mean & 0.689963 & 0.358651 & 20.635627   & 170.309794       & 223.489497       & 0.449108 & 0.930819 & 19.29     \\
             Standard deviation & 0.071367 & 0.125388 & 12.148875   & 112.633878       & 102.342995       & 0.127856 & 0.058487 &  1.97     \\
                         Median & 0.683189 & 0.361305 & 19.833554   & 134.222643       & 202.484306       & 0.428034 & 0.947978 & 19.50     \\
                        Q$_{1}$ & 0.626238 & 0.280797 & 13.580259   &  88.728917       & 171.597263       & 0.344535 & 0.913321 & 18.00     \\
                        Q$_{3}$ & 0.746521 & 0.463025 & 26.942293   & 280.103843       & 323.109765       & 0.525238 & 0.971689 & 20.25     \\
                            IQR & 0.120283 & 0.182228 & 13.362034   & 191.374926       & 151.512502       & 0.180702 & 0.058368 &  2.25     \\
                             OL & 0.445813 & 0.007455 & $-$6.462792 & $-$198.333472    & $-$55.671490     & 0.073482 & 0.825768 & 14.63     \\
                             OU & 0.926946 & 0.736367 & 46.985344   & 567.166232       & 550.378518       & 0.796291 & 1.059242 & 23.63     \\
          \hline
        \end{tabular}
        \label{knownatiras}
      \end{table*}
%
%

%
%
%
%

     Our results are robust, a representative set of control orbits (500) using the covariance matrix approach (see Section~2.2) give 
     consistent results within a few thousand years of our reference epoch. Dispersions resemble those in fig.~3 of 
     \citet{2019MNRAS.483L..37D}. It is however worth to mention that close encounters with Venus (see Fig.~\ref{evolution}, top panel, 
     left-hand side set), may induce significant orbital changes although the resulting orbit still remains within the Atira orbital realm.

     Atiras have an intrinsic interest within the context of planetary defense (see e.g. \citealt{2017JGRE..122..789M}) as they are major
     candidates to experience the so-called Red Baron dynamical scenario \citep{AD11} in which an asteroid approaches our planet from out of 
     the Sun's glare. A recent example of Red Baron scenario event that resulted in an impact is the one associated with the Chelyabinsk 
     superbolide \citep{2013MNRAS.436L..15D,2013Sci...342.1069P}, although in this case the parent body was not a member of the Atira class.

     Regarding the discoverability of these objects, {\it Gaia} \citep{2016A&A...595A...1G} may observe a number of objects at relatively 
     low solar elongations (see fig.~10 in \citealt{2018A&A...616A..13G}) and also scan at extreme inclinations with respect to the 
     ecliptic. Large Atiras and even Vatiras (apparent visual magnitude $<19$) at high inclination may be within reach of {\it Gaia}. 
     Vatiras may reach their aphelia more than twice per year, which increases the probability of accidental detection of such objects.
  
  \section{Conclusions}
     In this paper, we have explored the orbital evolution of 2019~AQ$_{3}$, which has the shortest aphelion of any known minor body, using 
     direct $N$-body simulations under the Newtonian and post-Newtonian approximations. Our conclusions are as follows: 
     \begin{enumerate}[(i)]
        \item The present-day orbital evolution of 2019~AQ$_{3}$ is relatively smooth as a result of being affected by the combined action 
              of the Earth--Moon system and Jupiter with Venus as a secondary perturber. The Lidov--Kozai mechanism is not at work here as 
              the argument of perihelion of 2019~AQ$_{3}$ does not librate, but circulates. This property is shared by several of the known 
              Atira-class asteroids and by some Atens. 
        \item Our calculations indicate that 2019~AQ$_{3}$ may have experienced brief stints as a Vatira in the relatively recent past and
              it may become a full-time Vatira in the near future after undergoing a close encounter with Venus, outlining a viable 
              dynamical pathway that may transform Atiras into Vatiras. Our calculations show that stable Vatiras can be trapped in a 
              Lidov--Kozai resonance that protects them from close encounters with Venus.
        \item Atira-class asteroid 2019~AQ$_{3}$ is an outlier within the known NEO orbits, but also an outlier (in terms of size) when 
              considering the predictions from a state-of-the-art orbit model. Among the known Atiras, it is a clear outlier in terms of 
              aphelion distance.
        \item The orbital evolution discussed here may keep Atiras switching between relatively close trajectories for extended periods of 
              time, but also transform Atiras into eventual Earth impactors coming from out of the Sun's glare.
        \item Although the value of the perihelion distance of 2019~AQ$_{3}$ can become as low as 0.17~au, the inclusion of post-Newtonian 
              terms in the numerical integrations does not significantly affect the overall orbital evolution.
        \item The dynamical evolution of the Atira-class asteroid 2018~JB$_{3}$ shows that Atens may evolve into Atiras/Vatiras and 
              therefore repopulate these dynamical classes when members are lost. 
     \end{enumerate}
     Atira-class asteroid 2019~AQ$_{3}$ is indeed a statistical outlier within the context of known NEAs, but it is also a rare and 
     important discovery that heralds a new dynamical class, that made of those NEAs following orbits entirely encompassed within that of 
     Venus: the Vatiras originally proposed by Greenstreet et al. (\citeyear{2012Icar..217..355G}). The evolution of the solar elongation of 
     2019~AQ$_{3}$ is similar to that of Venus with maximum elongation at about 47\degr. When observing from the ground, finding true 
     Vatiras requires the observation at elongations below 45\degr. This is at the edge of the minimum solar elongation scanned by the 
     spacecraft {\it Gaia} and it cannot be discarded that a true Vatira will be observed during the mission. A systematic exploration of 
     the Vatira orbital realm requires a well-designed, space-based mission.

  \section*{Acknowledgements}
     We thank the referee for his/her constructive reports and very helpful suggestions regarding the presentation of this paper and the 
     interpretation of our results, S.~J. Aarseth for providing the code used in this research, O. Vaduvescu, F. Roig, M.~N. De Pr{\'a}, and 
     S. Deen for comments, and A.~I. G\'omez de Castro for providing access to computing facilities. This work was partially supported by 
     the Spanish `Ministerio de Econom\'{\i}a y Competitividad' (MINECO) under grants ESP2015-68908-R and ESP2017-87813-R. In preparation of 
     this paper, we made use of the NASA Astrophysics Data System and the Minor Planet Center data server.

  \bsp
  \label{lastpage}

\begin{thebibliography}{99}
     \bibitem[\protect\citeauthoryear{Aarseth}{2003}]{2003gnbs.book.....A} Aarseth S.~J., 2003,
             Gravitational N-body Simulations.
             Cambridge Univ. Press, Cambridge, p.\ 27
     \bibitem[\protect\citeauthoryear{Adamo}{2011}]{AD11} Adamo D. R., 2011,
             Horizons Newsl., June 2011, p.\ 64
     \bibitem[\protect\citeauthoryear{Bellm \& Kulkarni}{2017}]{2017NatAs...1E..71B} Bellm E., Kulkarni S., 2017, 
             Nat. Astron., 1, 0071
     \bibitem[\protect\citeauthoryear{Benitez \& Gallardo}{2008}]{2008CeMDA.101..289B} Benitez F., Gallardo T., 2008, 
             Celest. Mech. Dyn. Astron., 101, 289
     \bibitem[\protect\citeauthoryear{Buzzi et al.}{2019}]{2019MPEC....A...88B} Buzzi L. et al., 2019,
             MPEC Circ., MPEC 2019-A88
     \bibitem[\protect\citeauthoryear{Chambers et al.}{2019}]{2019MPEC....A...97C} Chambers K. et al., 2019,
             MPEC Circ., MPEC 2019-A97
     \bibitem[\protect\citeauthoryear{de la Fuente Marcos \& de la Fuente Marcos}{2012}]{2012MNRAS.427..728D} de la Fuente Marcos C., 
             de la Fuente Marcos R., 2012, 
             MNRAS, 427, 728
     \bibitem[\protect\citeauthoryear{de la Fuente Marcos \& de la Fuente Marcos}{2013a}]{2013MNRAS.432..886D} de la Fuente Marcos C., 
             de la Fuente Marcos R., 2013a, 
             MNRAS, 432, 886
     \bibitem[\protect\citeauthoryear{de la Fuente Marcos \& de la Fuente Marcos}{2013b}]{2013MNRAS.436L..15D} de la Fuente Marcos C., 
             de la Fuente Marcos R., 2013b, 
             MNRAS, 436, L15
     \bibitem[\protect\citeauthoryear{de la Fuente Marcos \& de la Fuente Marcos}{2014}]{2014MNRAS.439.2970D} de la Fuente Marcos C., 
             de la Fuente Marcos R., 2014, 
             MNRAS, 439, 2970
     \bibitem[\protect\citeauthoryear{de la Fuente Marcos \& de la Fuente Marcos}{2015a}]{2015MNRAS.453.1288D} de la Fuente Marcos C., 
             de la Fuente Marcos R., 2015a, 
             MNRAS, 453, 1288
     \bibitem[\protect\citeauthoryear{de la Fuente Marcos \& de la Fuente Marcos}{2015b}]{2015A&A...580A.109D} de la Fuente Marcos C., 
             de la Fuente Marcos R., 2015b, 
             A\&A, 580, A109
     \bibitem[\protect\citeauthoryear{de la Fuente Marcos \& de la Fuente Marcos}{2018}]{2018RNAAS...2b..46D} de la Fuente Marcos C., 
             de la Fuente Marcos R., 2018, 
             Res. Notes AAS, 2, 46
     \bibitem[\protect\citeauthoryear{de la Fuente Marcos \& de la Fuente Marcos}{2019}]{2019MNRAS.483L..37D} de la Fuente Marcos C., 
             de la Fuente Marcos R., 2019, 
             MNRAS, 483, L37
     \bibitem[\protect\citeauthoryear{de la Fuente Marcos, de la Fuente Marcos \& Aarseth}{2015}]{2015MNRAS.446.1867D} de la Fuente Marcos C., 
             de la Fuente Marcos R., Aarseth S.~J., 2015, 
             MNRAS, 446, 1867
     \bibitem[\protect\citeauthoryear{Durda et al.}{2000}]{2000Icar..148..312D} Durda D.~D., Stern S.~A., Colwell W.~B., Parker J.~W., 
             Levison H.~F., Hassler D.~M., 2000, 
             Icarus, 148, 312
     \bibitem[\protect\citeauthoryear{Evans \& Tabachnik}{1999}]{1999Natur.399...41E} Evans N.~W., Tabachnik S., 1999, 
             Nature, 399, 41
     \bibitem[\protect\citeauthoryear{Evans \& Tabachnik}{2002}]{2002MNRAS.333L...1E} Evans N.~W., Tabachnik S.~A., 2002, 
             MNRAS, 333, L1
     \bibitem[\protect\citeauthoryear{Farinella et al.}{1994}]{1994Natur.371..315F} Farinella P., Froeschle Ch., Froeschle C., Gonczi R., Hahn G., 
             Morbidelli A., Valsecchi G.~B., 1994, 
             Nature, 371, 315
     \bibitem[\protect\citeauthoryear{Gaia Collaboration, Prusti et al.}{2016}]{2016A&A...595A...1G} Gaia Collaboration, Prusti T. et al., 2016,
             A\&A, 595, A1
     \bibitem[\protect\citeauthoryear{Gaia Collaboration, Spoto et al.}{2018}]{2018A&A...616A..13G} Gaia Collaboration, Spoto, F. et al., 2018, 
             A\&A, 616, A13
     \bibitem[\protect\citeauthoryear{Giorgini}{2011}]{2011jsrs.conf...87G} Giorgini J., 2011,
             in Capitaine N., ed.,
             Proceedings of the Journ\'ees 2010 ``Syst\`emes de r\'ef\'erence spatio-temporels'' (JSR2010):
             New challenges for reference systems and numerical standards in astronomy,
             Observatoire de Paris, Paris, p.\ 87
     \bibitem[\protect\citeauthoryear{Giorgini}{2015}]{2015IAUGA..2256293G} Giorgini J.~D., 2015,
             IAU General Assembly, Meeting \#29, 22, 2256293
     \bibitem[\protect\citeauthoryear{Giorgini \& Yeomans}{1999}]{GY99} Giorgini J. D., Yeomans D. K., 1999,
             On-Line System Provides Accurate Ephemeris and Related Data,
             NASA TECH BRIEFS, NPO-20416, p.\ 48
     \bibitem[\protect\citeauthoryear{Giorgini et al.}{1996}]{1996DPS....28.2504G} Giorgini J.~D. et al., 1996,
             BAAS, 28, 1158
     \bibitem[\protect\citeauthoryear{Giorgini, Chodas \& Yeomans}{2001}]{2001DPS....33.5813G} Giorgini J.~D., Chodas P.~W., Yeomans D.~K., 2001,
             BAAS, 33, 1562
     \bibitem[\protect\citeauthoryear{Granvik et al.}{2018}]{2018Icar..312..181G} Granvik M. et al., 2018,
             Icarus, 312, 181
     \bibitem[\protect\citeauthoryear{Greenstreet, Ngo \& Gladman}{2010}]{2010DPS....42.1309G} Greenstreet S., Ngo H., Gladman B., 2010, 
             AAS/Div. Planet. Sci. Meeting Abstr., 42, 13.09
     \bibitem[\protect\citeauthoryear{Greenstreet, Ngo \& Gladman}{2012}]{2012Icar..217..355G} Greenstreet S., Ngo H., Gladman B., 2012, 
             Icarus, 217, 355
     \bibitem[\protect\citeauthoryear{Hernandez \& Bertschinger}{2015}]{2015MNRAS.452.1934H} Hernandez D.~M., Bertschinger E., 2015, 
             MNRAS, 452, 1934
     \bibitem[\protect\citeauthoryear{Hernandez \& Bertschinger}{2018}]{2018MNRAS.475.5570H} Hernandez D.~M., Bertschinger E., 2018, 
             MNRAS, 475, 5570
     \bibitem[\protect\citeauthoryear{Kaiser}{2004}]{2004SPIE.5489...11K} Kaiser N., 2004, 
             SPIE, 5489, 11
     \bibitem[\protect\citeauthoryear{Kaiser et al.}{2004}]{2004AAS...204.9701K} Kaiser N., Pan-STARRS Project Team, 2004,
             BAAS, 36, 828
     \bibitem[\protect\citeauthoryear{Kozai}{1962}]{1962AJ.....67..591K} Kozai Y., 1962, 
             AJ, 67, 591
     \bibitem[\protect\citeauthoryear{Laskar et al.}{2011}]{2011A&A...532A..89L} Laskar J., Fienga A., Gastineau M., Manche H., 2011, 
             A\&A, 532, A89
     \bibitem[\protect\citeauthoryear{Libert \& Tsiganis}{2009}]{2009A&A...493..677L} Libert A.-S., Tsiganis K., 2009, 
             A\&A, 493, 677
     \bibitem[\protect\citeauthoryear{Lidov}{1962}]{1962P&SS....9..719L} Lidov M.~L., 1962, 
             Planet. Space Sci., 9, 719
     \bibitem[\protect\citeauthoryear{Mainzer}{2017}]{2017JGRE..122..789M} Mainzer A., 2017,
             JGRE, 122, 789
     \bibitem[\protect\citeauthoryear{Makino}{1991}]{1991ApJ...369..200M} Makino J., 1991,
             ApJ, 369, 200
     \bibitem[\protect\citeauthoryear{Masi}{2003}]{2003Icar..163..389M} Masi G., 2003, 
             Icarus, 163, 389
     \bibitem[\protect\citeauthoryear{Michel \& Thomas}{1996}]{1996A&A...307..310M} Michel P., Thomas F., 
             1996, A\&A, 307, 310
     \bibitem[\protect\citeauthoryear{Michel et al.}{2000}]{2000Icar..143..421M} Michel P., Zappal{\`a} V., Cellino A., Tanga P., 2000, 
             Icarus, 143, 421
     \bibitem[\protect\citeauthoryear{Milani \& Nobili}{1992}]{1992Natur.357..569M} Milani A., Nobili A.~M., 1992, 
             Nature, 357, 569
     \bibitem[\protect\citeauthoryear{Milani et al.}{1989}]{1989Icar...78..212M} Milani A., Carpino M., Hahn G., Nobili A.~M., 1989, 
             Icarus, 78, 212
     \bibitem[\protect\citeauthoryear{Murray \& Dermott}{1999}]{1999ssd..book.....M} Murray C.~D., Dermott S.~F., 1999,
             Solar System Dynamics,
             Cambridge Univ. Press, Cambridge, p.\ 316
     \bibitem[\protect\citeauthoryear{Naoz}{2016}]{2016ARA&A..54..441N} Naoz S., 2016, 
             ARA\&A, 54, 441
     \bibitem[\protect\citeauthoryear{Ohsawa et al.}{2019}]{2019MPEC....C...10O} Ohsawa R. et al., 2019,
             MPEC Circ., MPEC 2019-C10
     \bibitem[\protect\citeauthoryear{Popova et al.}{2013}]{2013Sci...342.1069P} Popova O.~P. et al., 2013, 
             Science, 342, 1069
     \bibitem[\protect\citeauthoryear{Ribeiro et al.}{2016}]{2016MNRAS.458.4471R} Ribeiro A.~O., Roig F., De Pr{\'a} M.~N., Carvano J.~M., 
             DeSouza S.~R., 2016, 
             MNRAS, 458, 4471
     \bibitem[\protect\citeauthoryear{Rivera-Valentin et al.}{2017}]{2017CBET.4347....1R} Rivera-Valentin E.~G., Taylor P.~A., Virkki A., 
             Aponte-Hernandez B., 2017,
             CBET, 4347
     \bibitem[\protect\citeauthoryear{Schumacher \& Gay}{2001}]{2001A&A...368.1108S} Schumacher G., Gay J., 2001, 
             A\&A, 368, 1108
     \bibitem[\protect\citeauthoryear{Sekhar et al.}{2017}]{2017MNRAS.468.1405S} Sekhar A., Asher D.~J., Werner S.~C., Vaubaillon J., Li G., 2017, 
             MNRAS, 468, 1405
     \bibitem[\protect\citeauthoryear{Sitarski}{1992}]{1992AJ....104.1226S} Sitarski G., 1992, 
             AJ, 104, 1226
     \bibitem[\protect\citeauthoryear{Smith et al.}{2014}]{2014SPIE.9147E..79S} Smith R.~M. et al., 2014, 
             SPIE, 9147, 914779
     \bibitem[\protect\citeauthoryear{Souami \& Souchay}{2012}]{2012A&A...543A.133S} Souami D., Souchay J., 2012, 
             A\&A, 543, A133
     \bibitem[\protect\citeauthoryear{Standish}{1998}]{ST98} Standish E.~M., 1998,
             JPL Planetary and Lunar Ephemerides, DE405/LE405,
             Interoffice Memo. 312.F-98-048, Jet Propulsion Laboratory, Pasadena, California
     \bibitem[\protect\citeauthoryear{Steffl et al.}{2013}]{2013Icar..223...48S} Steffl A.~J., Cunningham N.~J., Shinn A.~B., Durda D.~D., Stern S.~A., 2013, 
             Icarus, 223, 48
     \bibitem[\protect\citeauthoryear{Stern \& Durda}{2000}]{2000Icar..143..360S} Stern S.~A., Durda D.~D., 2000, 
             Icarus, 143, 360
     \bibitem[\protect\citeauthoryear{Tholen}{1998}]{1998MPBu...25...42T} Tholen D.~J., 1998, 
             MPBu, 25, 42
     \bibitem[\protect\citeauthoryear{Tholen \& Whiteley}{1998}]{1998DPS....30.1604T} Tholen D.~J., Whiteley R.~J., 1998, 
             AAS/Div. Planet. Sci. Meeting Abstr., 30, 16.04
     \bibitem[\protect\citeauthoryear{Tukey}{1977}]{1977eda..book.....T} Tukey J.~W., 1977,
             Exploratory Data Analysis.
             Addison-Wesley, Reading, MA
     \bibitem[\protect\citeauthoryear{Varadi, Runnegar \& Ghil}{2003}]{2003ApJ...592..620V} Varadi F., Runnegar B., Ghil M., 2003, 
             ApJ, 592, 620
     \bibitem[\protect\citeauthoryear{Vokrouhlick{\'y}, Farinella \& Bottke}{2000}]{2000Icar..148..147V} Vokrouhlick{\'y} D., Farinella P., 
             Bottke W.~F., 2000, 
             Icarus, 148, 147
     \bibitem[\protect\citeauthoryear{Wall \& Jenkins}{2012}]{2012psa..book.....W} Wall J.~V., Jenkins C.~R., 2012,
             Practical Statistics for Astronomers.
             Cambridge Univ. Press, Cambridge
     \bibitem[\protect\citeauthoryear{Warell, Karlsson \& Skogl{\"o}v}{2003}]{2003A&A...411..291W} Warell J., Karlsson O., Skogl{\"o}v E., 2003, 
             A\&A, 411, 291
     \bibitem[\protect\citeauthoryear{Whiteley \& Tholen}{1998}]{1998DPS....30.1603W} Whiteley R.~J., Tholen D.~J., 1998, 
             AAS/Div. Planet. Sci. Meeting Abstr., 30, 16.03
     \bibitem[\protect\citeauthoryear{Zavodny et al.}{2008}]{2008Icar..198..284Z} Zavodny M., Jedicke R., Beshore E.~C., Bernardi F., Larson S., 2008, 
             Icarus, 198, 284
  \end{thebibliography}
\end{document}